\documentclass[superscriptaddress,prl,twocolumn]{revtex4-1} 

\usepackage{comment}
\usepackage{graphicx}
\usepackage{float}
\usepackage{epstopdf}
\usepackage{color}
\usepackage{amsmath}
\usepackage{amssymb}
\usepackage{braket}
\usepackage{chngcntr}

\begin{document}

\title{Convergence to the asymptotic large deviation limit}

\author{Maxime Debiossac}
\affiliation{University of Vienna, Faculty of Physics, VCQ, Boltzmanngasse 5, A-1090 Vienna, Austria} 
\author{Nikolai Kiesel}
\affiliation{University of Vienna, Faculty of Physics, VCQ, Boltzmanngasse 5, A-1090 Vienna, Austria}
\author{Eric Lutz}
\affiliation{Institute for Theoretical Physics I, University of Stuttgart, D-70550 Stuttgart, Germany}

\begin{abstract} 
Large deviation theory offers a powerful and general statistical  framework to study the asymptotic dynamical properties of rare events. The application of the formalism to concrete experimental situations is, however,  often restricted by finite statistics. Data might not suffice to reach the asymptotic regime or  judge whether large deviation estimators converge at all.
  We here experimentally investigate the large deviation properties of the stochastic work and heat of a levitated nanoparticle subjected to nonequilibrium feedback control. This setting allows us to determine for each quantity the convergence domain of the large deviation estimators  using a criterion that does not  require the knowledge of the probability distribution. By extracting  both the asymptotic exponential decay and  the subexponential prefactors, we demonstrate that  singular prefactors  significantly restrict the convergence characteristics. Our results provide unique insight into the approach to the asymptotic large deviation limit and underscore the pivotal role of singular prefactors.  
\end{abstract}
\maketitle

Large deviation theory deals with  the probabilities  of exponentially rare fluctuations in stochastic systems. Such extreme events strongly deviate from  typical average  values. They hence evade  the law of large numbers and the central-limit theorem~\cite{ell85,deu89,dem98}. As a result, they are  in general  difficult to investigate both numerically and experimentally \cite{gia06,rag18}. Initiated by Cram\'er in the 1930s and further developed by Donsker and Varadhan  and by Freidlin and Wentzell in the 1970s, the theory of large deviations allows one to estimate the asymptotic  distributions of atypical events in the limit of a large scaling parameter. Examples  include the  distribution of  time-averaged  observables, such as  energy or particle currents flowing through a system, in the limit of long times~\cite{ell85,deu89,dem98}. Owing  to its versatility, the large deviation framework has found widespread applications in the analysis of random processes, from finance \cite{fri15}, statistics \cite{buc90} and engineering \cite{wei95}, to biology \cite{bre14} and physics \cite{vul14}. In this context, large deviation techniques have played an important role in the theoretical investigation of the fluctuating properties of small systems   in and out of equilibrium \cite{oon89,tou09}. Experimental studies of large deviation functions   in physical systems have  been presented in Refs.~\cite{bon09,kum11}.

Since empirical  data are always finite, a crucial issue that restricts the  practical applicability of the large deviation method is assessing the convergence towards the asymptotic regime \cite{roh15,nem17,hid17,whi18,rag20}. In particular, a nontrivial task is to determine the convergence region of  large deviation estimators from  available data for finite samples. Estimators might indeed converge slowly or not converge at all \cite{roh15,nem17,hid17,whi18,rag20}. Typical problems that have to be faced are the artificial linearization of the tails when the statistics is dominated by the largest value in the sample, and the fact that convergence is generally not uniform \cite{roh15,nem17,hid17,whi18,rag20}. Similar issues occur in the study of multifractals \cite{muz08,bac10}, glassy phase transitions \cite{ber11} and free energy estimators \cite{gor03,jar06,cog23}. Despite its central importance, the convergence  of estimators of large deviation functions  has not been investigated experimentally yet.

We here report a systematic experimental study of the convergence properties of the time-averaged heat and work in the context of stochastic thermodynamics \cite{sei12}. For the implementation, we use an optically levitated nanoparticle driven out of equilibrium with electric feedback control \cite{gie18,gie14,hoa18,deb20,gon21,deb22}.
 We  analyze both the convergence interval (related to the finite sample size) and the convergence time (related to the finite averaging time). To that end, we use a recently proposed, general convergence criterion, based on the evaluation of the standard error \cite{roh15}, that does not  require the knowledge of the probability distribution. Our experimental setup possesses  a number of unique features. First, in contrast to usual time series with fixed convergence characteristics, the convergence regions may be controlled using the feedback delay as a parameter. We identify values where only one of the two estimators is expected to converge, and others  where both (or none) of them are predicted to be convergent. Second, some large deviation properties, such as the asymptotic scaled cumulant generating functions of work and heat,  are analytically known for this system in the limit of high quality factors~\cite{ros17}, making a direct assessment of the accuracy of the estimation possible. Third, exploiting the high stability of our device, we measure a sufficiently large data sample to determine not only the asymptotic  exponential decay  of the distributions, but also the corresponding subexponential prefactors. These prefactors are subdominant in the long-time limit \cite{ell85,deu89,dem98}. However, our  results reveal that they play a decisive role in the approach to the asymptotic limit. We show that the presence of singularities in the prefactors of the moment generating function, which is the case of the stochastic heat, may dramatically affect the convergence rate and, at the same time, drastically restrict its convergence region.

\begin{figure}[t!]	\includegraphics[width=0.9\linewidth]{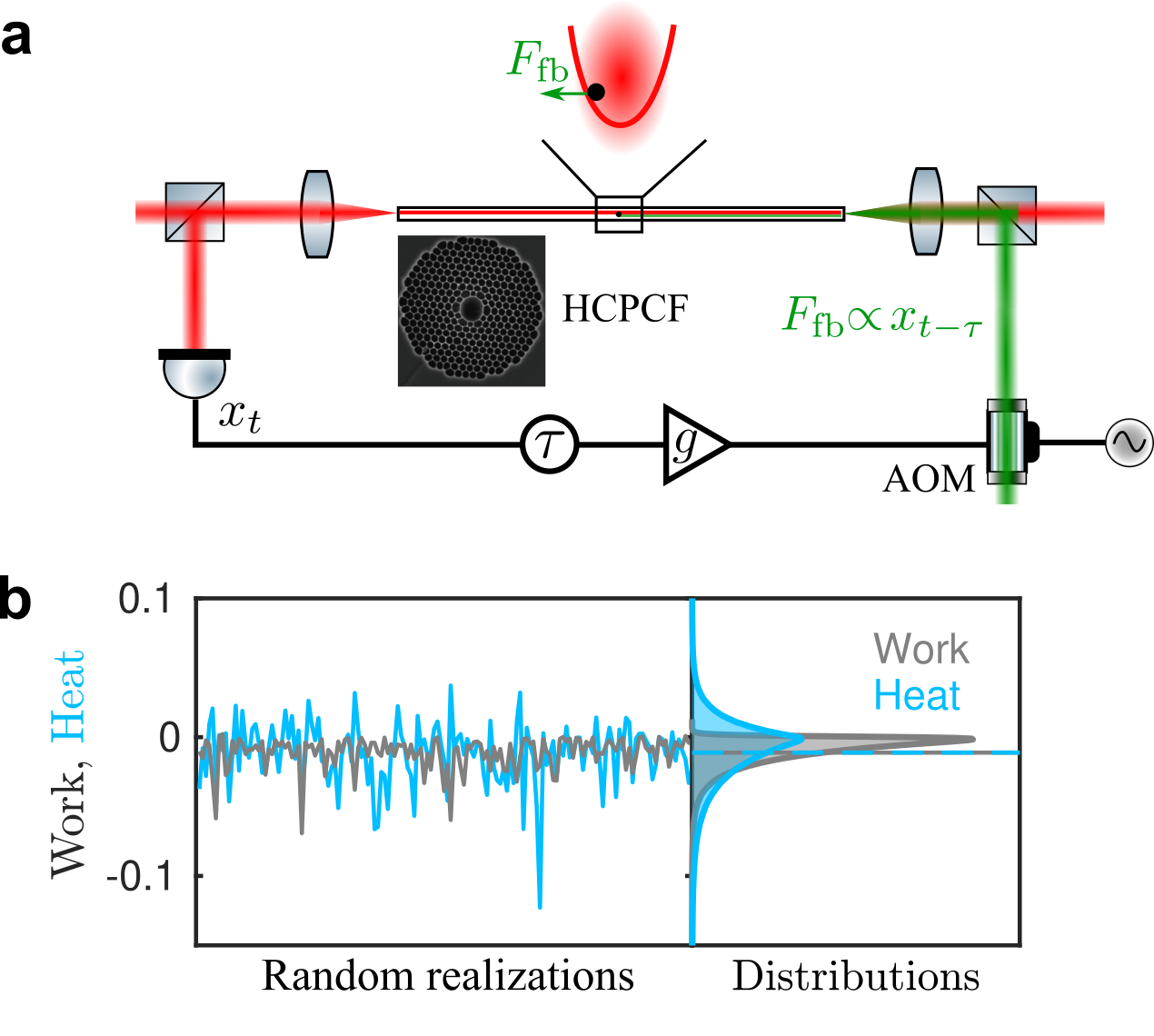}
	\caption{Experimental system. a) Schematic of the experimental setup consisting of a nanoparticle trapped in a harmonic potential inside a hollow core photonic crystal fiber (HCPCF). The particle is subjected to a delayed feedback force $F_\text{fb}\propto x_{t-\tau}$, with delay $\tau$  and  feedback gain $g$. b) Example of measured stochastic work (gray) and heat  (blue) realizations with their respective probability distributions, for delay $\tau=7.07$ and duration  $\mathcal{T}=Q_0$. The two distributions have the same mean (horizontal lines) but different tails.} 
	\label{fig:setup}
\end{figure}

\textit{Large-deviation estimators.}
Let us consider a random variable $\mathcal{A}$ and its average over a time interval $\mathcal{T}$, $a=\int_0^\mathcal{T}dt \mathcal{A}(t)/\mathcal{T}$. In the following, $\mathcal{A}$ will be either the stochastic work  or the stochastic heat  of the levitated nanoparticle. The central quantities of large deviation theory are the probability distribution $P(a)$, the rate function $I(a)$, and the scaled cumulant generating function $\mu_\mathcal{A}(\lambda)$ \cite{ell85,deu89,dem98}. The large deviation principle states that, for  large time $\mathcal{T}$, the probability $P(a)$  decays exponentially  with rate $I(a)$,  $P(a)\sim e^{-I(a)\mathcal{T}}$. To evaluate the rate function, it is often convenient to introduce the scaled cumulant generating function, $\mu_\mathcal{A}(\lambda)=\lim_{\mathcal{T}\rightarrow \infty}\ln \langle e^{-\lambda\mathcal{A}}\rangle/\mathcal{T}$, where $\langle .\rangle$ denotes the ensemble average. When  $\mu_\mathcal{A}(\lambda)$ is differentiable, the rate function   follows  as the Legendre transform $I(a) = -\lambda_* a - \mu_\mathcal{A}(\lambda_*)$  with $\lambda_*(a)$ being the root of $ \mu'_\mathcal{A}(\lambda_*) = - a$ \cite{ell85,deu89,dem98}.

Estimating the large deviation properties from   data is usually a delicate task due to the finite sample size. We employ a block averaging method that divides  the total length, $\mathcal{T}_\text{tot}=N\mathcal{T}$, of the time series into $N$ blocks of duration $\mathcal{T}$ \cite{ytreberg2004,duffy05}. Finite data  imposes constraints on the number $N$ of trajectories  that can be used for ensemble average calculations and on the length $\mathcal{T}$ of  the trajectories. The statistical   estimators of the scaled cumulant generating function and  of the rate function are respectively given by $\mu_\mathcal{A}(\lambda,\mathcal{T},N)=(1/\mathcal{T})\ln(\sum_{i=1}^{N} e^{-\lambda\mathcal{A}_i}/N)$ and $I_\mathcal{A}(a,\mathcal{T},N)=-\lambda a_\mathcal{A}-\mu_\mathcal{A}(\lambda,\mathcal{T},N)$, with the average $a_\mathcal{A}=-(1/\mathcal{T})(\sum_{i=1}^{N} \mathcal{A}_i e^{-\lambda \mathcal{A}_i}/\sum_{i=1}^{N} e^{-\lambda \mathcal{A}_i})$~\cite{roh15,nem17,hid17,whi18,rag20}. These estimators are expected to asymptotically converge to their large deviation limits, $\mu_\mathcal{A}(\lambda)$ and $I(a)$, for sufficiently large $N$ and $\mathcal{T}$~\cite{roh15,nem17,hid17,whi18,rag20}.

\begin{figure*}[t]
	\includegraphics[width=1.03\linewidth]{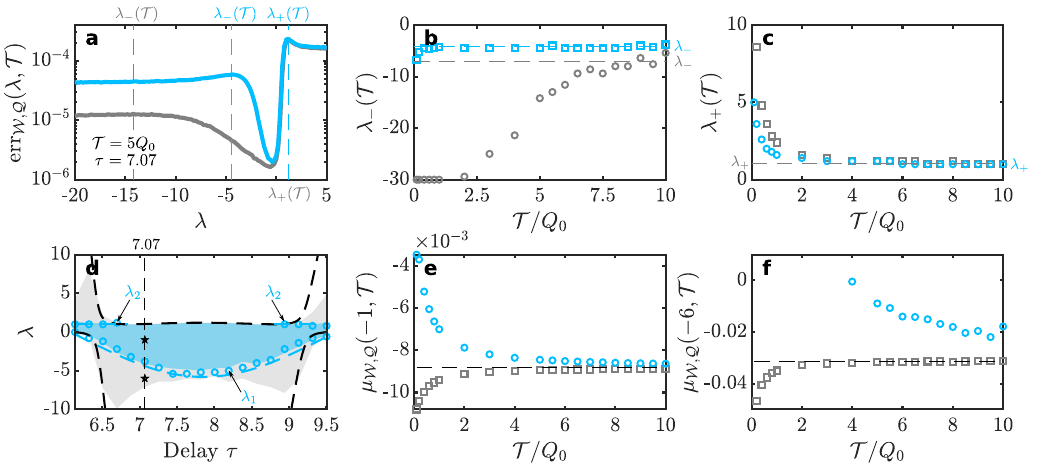}
	\caption{Convergence regions. a) Standard error  $\text{err}_\mathcal{W,Q}(\lambda,\mathcal{T})$ of work (gray) and heat (blue), for duration $\mathcal{T}=5Q_0$ and delay $\tau=7.07$. The maxima at $\lambda_\pm(\mathcal{T})$ indicate the onset of the linearization of the tails when the statistics is dominated by the largest element of the sample.  b)-c) The boundaries $\lambda_\pm$ of the convergence domains  are associated with the (averaged) values of the plateaus of $\lambda_\pm(\mathcal{T})$ for large $\mathcal{T}$. d) Convergence regions of the scaled cumulant generating functions of work and heat for all values of the delay.  The boundaries for work (and partially for heat) correspond to the analytically known asymptotic range of definition (black dashed line). The boundaries for heat are restricted by singularities of the prefactor: good agreement is seen between the measured singularities (blue open circle) and the analytic expressions, $\lambda_1= -1-2g\sin \tau$ and $\lambda_2=1$, derived in the high $Q_0$ approximation (blue dashed line). e) The scaled cumulant generating functions $\mu_\mathcal{W,Q}(\lambda,\mathcal{T})$ of work and  heat both converge for $\lambda= -1$, whereas f) only the one of work converges for $\lambda= -6$. Statistical error bars are discussed in the Supplementary Information.
			\label{Fig:linearization}}  
\end{figure*}

\begin{figure*}[t]	\includegraphics[width=1.01\linewidth]{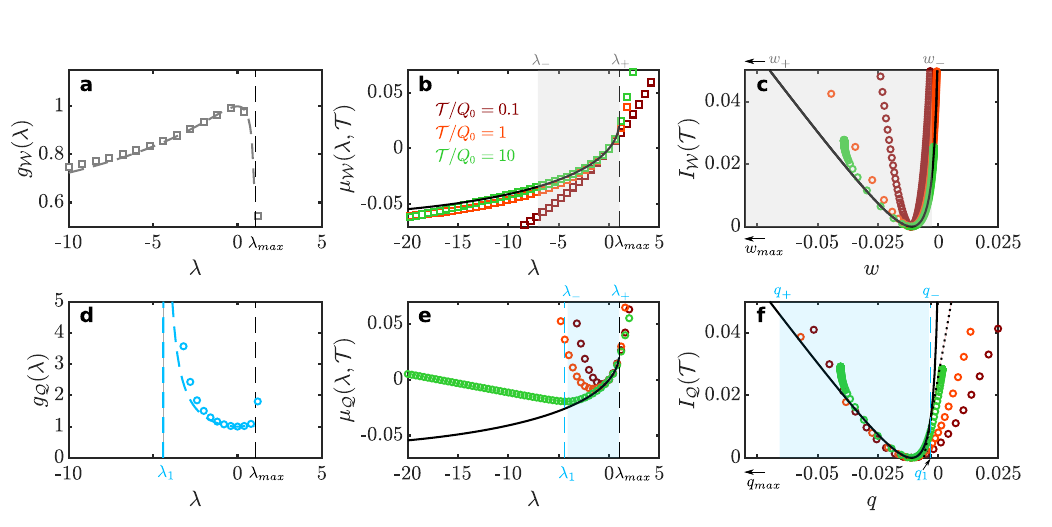}
	\caption{Singular prefactors and large deviation estimators. a) and d) Experimentally determined subexponential prefactors $g_\mathcal{W,Q}(\lambda)$ of work and heat (symbols) for $\tau=7.07$ and analytical results derived in the high-$Q_0$ approximation (dashed lines). The prefactor of heat exhibits a singularity at $\lambda = \lambda_1$. The edge of the asymptotic theoretical range of definition is denoted by $\lambda_\text{max}$. b) and e) Estimators of the scaled cumulant generating functions $\mu_\mathcal{W,Q}(\lambda,\mathcal{T})$ of work and  heat for various durations $\mathcal{T}$ (symbols) and asymptotic theoretical prediction (black solid lines). The convergence domains are represented by the (gray and blue) areas. The pole of the prefactor  restricts   the convergence interval and the convergence time of the  heat. c) and f) Estimators of the rate functions $I_\mathcal{W,Q}(\mathcal{T})$ of work and  heat for various $\mathcal{T}$ (symbols). Whereas $I_\mathcal{W}(\mathcal{T})$ approaches the asymptotic limit (black solid line) for large $\mathcal{T}$, $I_\mathcal{Q}(\mathcal{T})$ deviates from it due to the linearization induced by the singularity (dotted line). Statistical error bars are discussed in the Supplementary Information.
}
	\label{Fig:prefactors}
\end{figure*}

 \textit{Experimental system.} Levitodynamics has played a key role in the investigation of the stochastic thermodynamics of small underdamped systems \cite{gie18,gie14,hoa18,deb20,gon21,deb22}. In our experiment, we trap a levitated silica nanoparticle (295 nm diameter)  at an intensity maximum of a standing wave formed by two counterpropagating laser beams ($\lambda_0=1064$ nm) inside a hollow-core photonic crystal fiber (HCPCF) (Fig.~\ref{fig:setup}a) \cite{gra16}. The particle is characterized by a quality factor $Q_0=\Omega_0/\Gamma_0=50.2$, where $\Omega_0/2\pi=297.7$ kHz is the resonance frequency and $\Gamma_0/2\pi=5.93$  kHz is the damping due to the gas  inside the HCPCF (temperature $T=293$ K). We detect the position $x_t$ of the particle along the fiber axis  with  an interferometric readout of the light scattered by the particle, with a  sensitivity of 2 pm/$\sqrt{\text{Hz}}$~\cite{gra16}. A feedback force, $F_{\text{fb}}= -g (m \Gamma_0 \Omega_0) x_{t-\tau}$, is applied to the nanoparticle (of mass $m$) via radiation pressure exerted by an additional laser beam (green beam in Fig.~\ref{fig:setup}a).  We vary the delay  $\tau$  using a field-programmable gate array. The feedback loop has a gain $g=2.4$ and an internal minimum delay of $3~\mu$s. 
We select 19 different  values of the delay, belonging to a stability region of the feedback loop, and record for each  a long trajectory of duration $\mathcal{T}_\text{tot}=1000\:$s with a sampling rate of $5$ MHz. This correspond to 5 hours total data acquisition time, during which the relative uncertainty on the damping rate and feedback gain stays smaller than 5\%. 

The stochastic work along a trajectory of duration $\mathcal{T}$ is defined as $\mathcal{\beta W}=(2g/Q_0^2)\int_0^\mathcal{T}dtx_{t-\tau}\circ v_t$, in dimensionless units, with the Stratonovich-type product $\circ$ \cite{sei12} and the inverse temperature $\beta$. According to the first law, the random heat is given by $\mathcal{Q}=\mathcal{W}-\Delta\mathcal{U}$, where $\Delta\mathcal{U}=(x^2_\mathcal{T}-x^2_0+v^2_\mathcal{T}-v^2_0)/Q_0$ is the change in internal energy. Work and heat thus only differ through the temporal boundary term $\Delta\mathcal{U}$, and have equal mean. However, their fluctuation and large deviation features are  different. Figure~1b shows examples of measured work (gray) and heat (blue) realizations for a delay $\tau=7.07$, as well as the corresponding distributions. The tails of the heat are much broader than those of the work because of the additional fluctuations of the boundary term.

{\textit{Convergence regions.} The convergence domain of the statistical estimators may be determined by inspecting the linearization phenomenon of the scaled cumulant generating function~\cite{roh15,nem17,hid17,whi18,rag20}. This effect, which depends on the values of $\lambda$ (nonuniform convergence), occurs when the sum $\sum_{i=1}^N e^{-\lambda\mathcal{A}_i}$ is dominated by its largest element. This translates into a finite convergence interval $\Delta\lambda=\lambda_+-\lambda_-$ delimited by $\lambda_-<0$ and $\lambda_+>0$. The values $\lambda_\pm$ depend in general on $\mathcal{T}$ and $N$. In order to account for possible correlations between data points, we  group (reshuffled) data into $N_b$ (asymptotically) independent blocks of size $N/N_b$ \cite{ytreberg2004,duffy05}. 
We choose a large value of  $N=5\times 10^6$ for the number of trajectories and set $N_b=1000$. Following the theoretical suggestion of Ref.~\cite{roh15}, we estimate  $\lambda_{\pm}$ by determining the (positive and negative) maxima of the standard error of $a_\mathcal{A}$ in each block, $\text{err}_\mathcal{A}(\lambda,\mathcal{T})=\text{std}(a_\mathcal{A})/\sqrt{N_b}$ where $\text{std}$ is the standard deviation, for large $\mathcal{T}$, since these  maxima are related to the onset of  linearization (Supplementary Information).
  Figure~\ref{Fig:linearization}a displays $\text{err}_\mathcal{A}(\lambda,\mathcal{T})$ for work and heat, as a function of $\lambda$,  for  time $\mathcal{T}/Q_0=5$ and delay  $\tau= 7.07$ (other values are presented in the Supplementary Information). We associate  $\lambda_\pm$ with the (averaged) plateaus of $\lambda_\pm(\mathcal{T}) $ seen in Figs.~\ref{Fig:linearization}b-c,
 as a function of $\mathcal{T}$, for  $\mathcal{T}>8Q_0$ (dashed lines). The values of   $\lambda_\pm$ do not depend on $N$, for $N$ large enough (Supplementary Information). The respective standard errors for the scaled cumulant generating functions $\mu_\mathcal{A}$  for  work and heat are analyzed in the Supplementary Information.

The respective convergence regions 	 of the  estimators of the scaled cumulant generating function of work and heat are obtained by repeating the same procedure for all delays (gray and blue shaded areas in Fig.~\ref{Fig:linearization}d).  The boundaries $\Delta\lambda$ are set by the delay.
We observe three different regimes depending on the value of $\lambda$: (i) a regime where the statistical estimators of both work and heat are expected to converge (overlapping gray and blue areas), (ii) a domain where only the estimator of work is expected to converge (non-overlapped gray area) and (iii) a region where none of them is predicted to be convergent (white area). Interestingly,  the   boundaries $\lambda_\pm$  for the  stochastic work coincide  precisely with  the analytically  known asymptotic range of definition of the scaled cumulant generating function, where the latter is real (dashed black lines) \cite{ros17} (Supplementary Information); the cuts seen at the top and at the bottom are due to finite statistics. A similar effect  has   been noticed theoretically in another context in Ref.~\cite{roh15}. These findings demonstrate the  power of the simple convergence criterion based on the standard error, a scheme that does not require the knowledge of the  distribution of the random variable considered. Surprisingly, even though work and heat are equal on average, the predicted convergence region for the stochastic heat is markedly  smaller than that for work. It  also appears to be systematically more restricted than the analytically  known asymptotic range of definition of the scaled cumulant generating function   (same dashed black lines as for the work) \cite{ros17}.

\textit{Singular prefactors.} In order to elucidate this difference between work and heat, we  examine the subexponential prefactors, $g_\mathcal{A}(\lambda)$, of the corresponding moment generating functions, defined as $\langle e^{-\lambda \mathcal{A}}\rangle \sim  g_\mathcal{A}(\lambda) e^{\mu_\mathcal{A}(\lambda) /\mathcal{T}}$, for large $\mathcal{T}$. Theoretical computations of the  prefactors are  difficult in general. For the problem at hand, analytical  formulas for $g_\mathcal{A}(\lambda)$ are unknown for  generic values of $\lambda$ and $\tau$, except in the small-$\tau$ or high-$Q_0$ (Markovian) limits \cite{ros17}. Likewise, extracting prefactors from experimental data is  usually hard, since they correspond to very small deviations  for small trajectory length. In order to do so, we fit the functions $\mu_\mathcal{A}(\lambda,\mathcal{T})$ in the range $1<\mathcal{T}/Q_0<10$ for each $\lambda$,  using the expression $A+B/\mathcal{T}+C/\mathcal{T}^2$ (Supplementary Information). The prefactors are accordingly given by  $g_\mathcal{A}(\lambda)=e^B$. Figures 3a,d present the experimentally determined prefactors for work and heat as a function of $\lambda$ for the delay  $\tau= 7.07$ (other values are again presented in the Supplementary Information). Whereas the prefactor $g_\mathcal{W}(\lambda)$ for work is finite for all $\lambda$ (Fig.~\ref{Fig:prefactors}a), the prefactor $g_\mathcal{Q}(\lambda)$ for heat exhibits a singularity around $\lambda=\lambda_1\simeq -4.4$ (Fig.~\ref{Fig:prefactors}d). Such a diverging behavior is due  to rare but large fluctuations of the boundary term $\Delta \mathcal{U}$.  In both instances, we have excellent  agreement  between data and  the analytical approximations (gray and blue dashed lines) obtained for high quality factors.

The locations of the experimentally evaluated singularities of the heat prefactor $g_\mathcal{Q}(\lambda)$ for all implemented delays are shown in Fig.~\ref{Fig:linearization}d (blue open circle) together with the corresponding analytical high-$Q_0$ expressions (blue dashed lines). They perfectly match the predicted convergence region based on the standard-error criterion. We may thus conclude that, although subexponential prefactors of the scaled cumulant generating function become irrelevant in the asymptotic large-deviation limit,  their singularities strongly restrict the convergence domain of the corresponding statistical estimator.

\textit{Convergence time.} We next analyze the influence  of a singular prefactor on the convergence time of the statistical estimators of the scaled cumulant generating function $\mu_\mathcal{A}(\lambda,\mathcal{T})$ and of the rate function $I_\mathcal{A}(a,\mathcal{T})$. Figures 3b,e exhibit the estimator of the scaled cumulant generating function $\mu_\mathcal{W,Q}(\lambda,\mathcal{T})$ of work and heat as a function of $\lambda$, for different trajectory lengths. We notice that $\mu_\mathcal{W}(\lambda,\mathcal{T})$ (symbols) quickly converges to the analytically known asymptotic limit (black line) within the convergence region (grey dashed area). The linearization of the tail, and the corresponding deviation from the theoretical asymptotic limit, are clearly visible outside the convergence domain. By contrast, the presence of the singularity at $\lambda = \lambda_1$ significantly slows down the convergence of $\mu_\mathcal{Q}(\lambda,\mathcal{T})$ close to $\lambda_1$ at short times. For longer times, as the asymptotic regime is approached, the  effect of the divergence of the prefactor is suppressed, as expected. However, linearization occurs at smaller values of $\lambda$ compared to  the work, as discussed above, even before the asymptotic regime can be  reached for all $\lambda$. Singular prefactors therefore reduce both the convergence interval and the converge time of the statistical estimator of the scaled cumulant generating function. Figures 3b,e additionally highlight the danger of evaluating statistical estimators without determining the convergence domain: not only can  estimators depart from the asymptotic result (as seen on the left-hand side of the figures), they can also be computed from data for parameters where a scaled cumulant generating does not exist in the asymptotic limit (as seen on the right-hand side).

The statistical estimators of the rate functions $I_\mathcal{A}(a,\mathcal{T})$ for work and heat are shown in Figs.~3c,f. The boundaries $a_\pm$ of the respective shaded areas are defined through $\mu'_\mathcal{A}(\lambda_\pm) = -a_\pm$. The estimator $I_\mathcal{W}(w,\mathcal{T})$ of  work (symbols) approaches the known analytic asymptotic result $I(w)$ (black solid line) as the length of the trajectory increases.  The situation is again different for  heat. The corresponding exact expression for the rate function  is    not analytically known due to the singularity. The pole of the prefactor  indeed modifies the large deviation function $I(q)$, which is no longer correctly described by the Legendre transform of $\mu_\mathcal{Q}(\lambda)$ (black solid line) close to the singularity $q_1$. The leading contribution to the rate function comes essentially  from the pole. This leads to an exponential tail of the distribution, or, equivalently, to a linear large deviation function for $q>q_1$, whose slope is determined by the pole (black dotted line).


{\textit{Conclusions.}
Large deviation estimators are only useful when they converge to their -- usually unknown -- asymptotic limit. The convergence region and the converge rate are, however, equally unknown in general. We have experimentally investigated these important issues using a highly stable levitodynamic system subjected to feedback control. We have implemented, for the first time, a simple and reliable convergence criterion based on the standard error, that does not require the knowledge of the probability distribution. We have demonstrated that this criterion is  able to identify the asymptotic range of definition of the scaled cumulant generating function, as well as the presence of singular prefactors, which we could independently detect. Such divergences are known to occur in many linear systems, both for heat \cite{zon03,vis06,bai06,noh12} and work \cite{far02,far04,sab11,sab12}, as well as in nonlinear systems \cite{pug06,har06,nem12}. We have shown that they restrict both the convergence interval and the convergence time. These findings highlight the critical role of singular prefactors in the approach to the asymptotic large deviation limit. 

\textit{Acknowledgments.}
 We  acknowledge financial support from the German Science Foundation (DFG) (Project FOR 2724) and  the Austrian Science Fund (FWF) (Project Y 952-N36, START). We also thank Martin Luc Rosinberg for providing  the analytical results for the large-$Q$ expressions of the prefactors.

\clearpage
\pagebreak
\widetext

\setcounter{equation}{0}
\setcounter{figure}{0}
\setcounter{table}{0}
\setcounter{page}{1}
\makeatletter
\renewcommand{\theequation}{S\arabic{equation}}
\renewcommand{\thefigure}{S\arabic{figure}}
\renewcommand{\bibnumfmt}[1]{[S#1]}
\renewcommand{\citenumfont}[1]{S#1}

\newpage 
\begin{center}
\vskip0.5cm
{\large \bf Supplementary Information: Convergence to the asymptotic large deviation limit}
\end{center}


\section{I. Theoretical results}
\subsection{a. Equation of motion} 
The dynamics of the nanoparticle is well described by an underdamped Langevin equation of the form
\begin{align}
\ddot{x}_t+ \Gamma_0 \dot{x}_t+ \Omega_0^2 x_t- g \Gamma_0 \Omega_0 x_{t-\tau}=\sqrt{\frac{2\Gamma_0 k_BT}{m}}\xi_t,
\label{eq:langevin}
\end{align}  
where  $m$ is the mass, $\Gamma_0$ the damping coefficient,  $\Omega_0$ the resonance frequency and $\xi(t)$  a delta-correlated Gaussian thermal noise with unit variance. The linear delayed feedback is applied via the force  $F_{\text{fb}}= -g (m \Gamma_0 \Omega_0) x_{t-\tau}$.  Equation~\eqref{eq:langevin} may be expressed in a dimensionless form using $\Omega_0^{-1}$ and $x_0=(1/\Omega_0^2)\sqrt{2\Gamma_0 k_BT/m}$ as respective units of time and position. The particle dynamics  is then  characterized by a set of dimensionless parameters $(g,Q_0,\tau)$ \cite{S_ros17}. The length ${\cal T}$ of the analyzed trajectories may be further expressed in units of the quality factor $Q_0$.

\subsection{b. Scaled cumulant generating functions and subexponential prefactors}
In the asymptotic limit, the moment generating function, $Z_\mathcal{A}(\lambda,\mathcal{T})=\langle e^{-\lambda \mathcal{A}}\rangle$, behaves as $Z_\mathcal{A}(\lambda,\mathcal{T}) \sim g_\mathcal{A}(\lambda) e^{\mu_\mathcal{A}(\lambda) \mathcal{T}}$.
Assuming that the boundary term $\Delta \mathcal{U}$ plays no role in the long-time limit, the scaled cumulant generating function $\mu(\lambda) = \mu_{\mathcal{W,Q}}(\lambda)$ can be evaluated analytically via Fourier transformation \cite{S_ros17}:
\begin{equation}
\mu(\lambda)=-\frac{1}{2\pi}\int_0^\infty d\omega\ln\left[1-4\lambda g/Q_0^2\omega\sin(\omega\tau)|\chi(\omega)|^2\right],
\label{Eq:mu_th}
\end{equation}
where $\chi(\omega)=(-\omega^2-i\omega/Q_0+1-g/Q_0e^{i\omega\tau})^{-1}$ is the mechanical susceptibility. The range of definition of $\mu(\lambda)$ is given by $[\lambda_\text{min},\lambda_\text{max}]$, where $\lambda_\text{min}$ and $\lambda_\text{max}$ are defined such that the argument of the logarithm in Eq.~\eqref{Eq:mu_th} stays positive, ensuring that $\mu(\lambda)$ is real.

The analytical expressions of the subexponential prefactors $g_\mathcal{A}(\lambda)$ are not known  for generic values of $\lambda$ and $\tau$. In the high $Q_0$ limit (corresponding to the Markovian approximation), one finds using the methods of Ref.~\cite{S_ros17}
\begin{equation}
    g_\mathcal{W}(\lambda)=\frac{4(1+g\sin\tau)\Tilde{g}(\lambda,\tau)}{[1+g\sin\tau+\Tilde{g}(\lambda,\tau)]^2},
\end{equation}

\begin{equation}
    g_\mathcal{Q}(\lambda)=\frac{4(1+g\sin\tau)\Tilde{g}(\lambda,\tau)}{[1+g\sin\tau+\Tilde{g}(\lambda,\tau)]^2-4\lambda^2},
    \label{Eq:prefactor}
\end{equation}
with  $\Tilde{g}(\lambda,\tau)=\sqrt{(1+g\sin\tau)^2-4g\lambda\sin\tau}$. The prefactor of work, $g_\mathcal{W}(\lambda)$, does not exhibit any singularity.  However, the prefactor of  heat, $g_\mathcal{Q}(\lambda)$,  diverges  at $\lambda_2=1$ (in region I) and at $\lambda_1=-1-2g\sin\tau$ (in regions I and II) (Fig.~S1) .

\subsection{c. Large deviation functions}
The rate function of work $ I_\mathcal{W}(w)$ can be computed as the Legendre transform $ I_\mathcal{W} (w)= -\lambda_* w - \mu(\lambda_*)$ with $\mu'(\lambda_*) + w =0$. However, due to the singularities of the prefactor, the rate function of heat  $ I_\mathcal{Q}(q)$ is not known analytically. Only the tails of the rate function can be calculated. They are given for $q<q_2$ and $q>q_1$ by \cite{S_ros17}
\begin{equation}
    I_\mathcal{Q}=-\lambda_i q-\mu(\lambda_i),
\end{equation}
where the $\lambda_i$ are the positions of the singularities of the prefactor $g_\mathcal{Q}(\lambda)$ that correspond to the values $q_i$.

\section{II. Additional experimental results}
\subsection{a. Conjugate dynamics}

The system exhibits a dynamics that depends on the delay (Fig.~\ref{Fig:domain}). Depending on the existence of a steady state for a conjugate dynamics, two distinct regions with different physical properties can be identified~\cite{S_ros17}.  In region II ($6.85<\tau<8.80$), a conjugate dynamics of the system may be introduced  by changing the sign of the friction coefficient ($\Gamma_0\rightarrow-\Gamma_0$) in the Langevin equation~\cite{S_ros17}. In this region, we choose two delays to probe the effect of the divergence of the prefactors at $\lambda=\lambda_1$: $\tau=7.07$ presented in the main text and $\tau=8.01$ shown here. In region I ($6.00<\tau<6.85$ and $8.80<\tau<9.66$), a different conjugate dynamics may be introduced by inverting the sign of the delay in the Langevin equation. We choose a delay $\tau=6.13$ to sample the divergence of the prefactor at $\lambda=\lambda_2=1$.    

\begin{figure*}[h]	\includegraphics[width=0.55\linewidth]{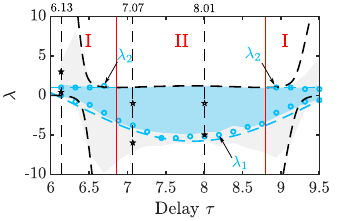}
	\caption{Different dynamics of the system. In region I, the heat prefactor exhibits a singularity at $\lambda = \lambda_1$ and $\lambda =\lambda_2$, whereas in region II, it displays  a singularity of $\lambda = \lambda_1$. The choice of the three delays investigated in detail, $\tau = 6.13$, $\tau = 7.07$,  and $\tau = 8.01$,  are indicated by vertical dashed lines. For each of these delays, we choose two values of $\lambda$ (black stars) to analyze the convergence of the statistical estimators as a function of time.}
	\label{Fig:domain}
\end{figure*}

\subsection{b. Fit of the prefactors of the moment generating function}
We determine the prefactors of the moment generating function by fitting the estimators of the scaled generating function for work and heat,  $\mu_\mathcal{A}(\lambda, \mathcal{T}) = \mu_\mathcal{A}(\lambda) + \ln g_\mathcal{A}(\lambda)/\mathcal{T}$, with   the expression $A+B/\mathcal{T}+C/\mathcal{T}^2$. The fit for the three delays $\tau = 6.13$, $\tau = 7.07$ and $\tau = 8.01$ are shown in Fig.~\ref{Fig:fit}.

\begin{figure*}[h]	\includegraphics[width=0.95\linewidth]{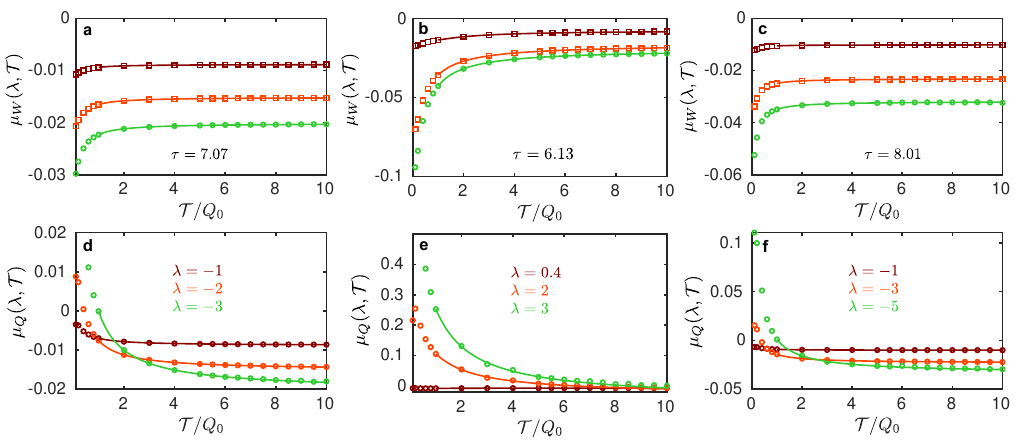}
	\caption{Fit of the scaled cumulant generating function for work (a,b,c) and heat (d,e,f) for various delays  $\tau=7.07$ (a,d), $\tau=6.13$ (b,e) and $\tau=8.01$ (c,f), and different values of $\lambda=(-1,-2,-3)$ for panels (a,d), $\lambda=(0.4,2,3)$ for panels (b,e) and $\lambda=(-1,-3,-5)$ for panels (c,f). The fit is performed in the range $\mathcal{T}/Q_0=1$ and $\mathcal{T}/Q_0=10$.
	}
	\label{Fig:fit}
\end{figure*}

\subsection{c. Delay $\tau=6.13$}

We show here experimental data for the delay $\tau=6.13$ (region I) following the same presentation as in the main text. In Fig.~\ref{Fig:ldf_tau1}, we have chosen the values of $\lambda=0.4$ and $\lambda=3$ (marked by a star in  Fig.~\ref{Fig:domain}). The deviations between data and analytical results seen in Fig.~\ref{Fig:ldf_tau1}a and Fig.~\ref{Fig:ldf_tau1}d are due to the breakdown of the high-$Q_0$ approximation for some values of the delay.

\begin{figure*}[h!]	\includegraphics[width=\linewidth]{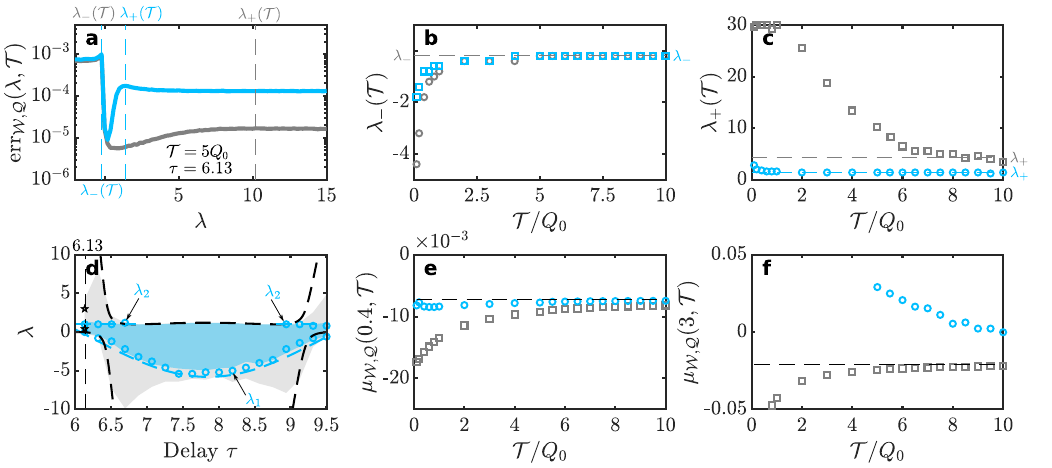}
	\caption{Same plots as in Fig.~2 of the main text for the delay $\tau=6.13$.
	}
	\label{Fig:linear_tau1}
\end{figure*}

\begin{figure*}[h!]	\includegraphics[width=\linewidth]{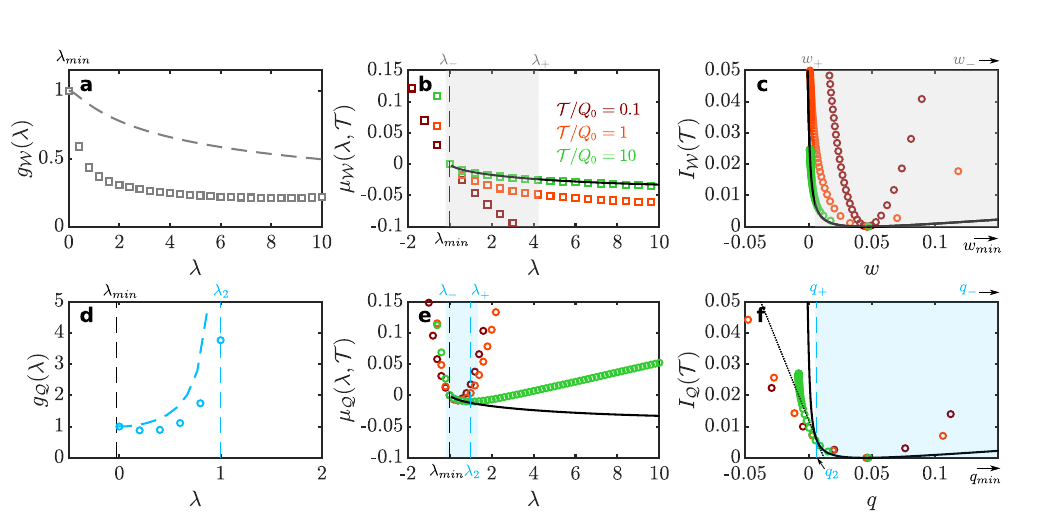}
	\caption{Same plots as in Fig.~3 of the main text for the delay $\tau=6.13$.
	}
	\label{Fig:ldf_tau1}
\end{figure*}

\subsection{c. Delay $\tau=8.01$}

We show here experimental data for the delay $\tau=8.01$ (region II) following the same presentation as in the main text. In Fig.~\ref{Fig:ldf_tau2}, we have chosen the values of $\lambda=-1$ and $\lambda=-5$ (marked by a star in Fig.~\ref{Fig:domain}).

\begin{figure*}[h!]	\includegraphics[width=\linewidth]{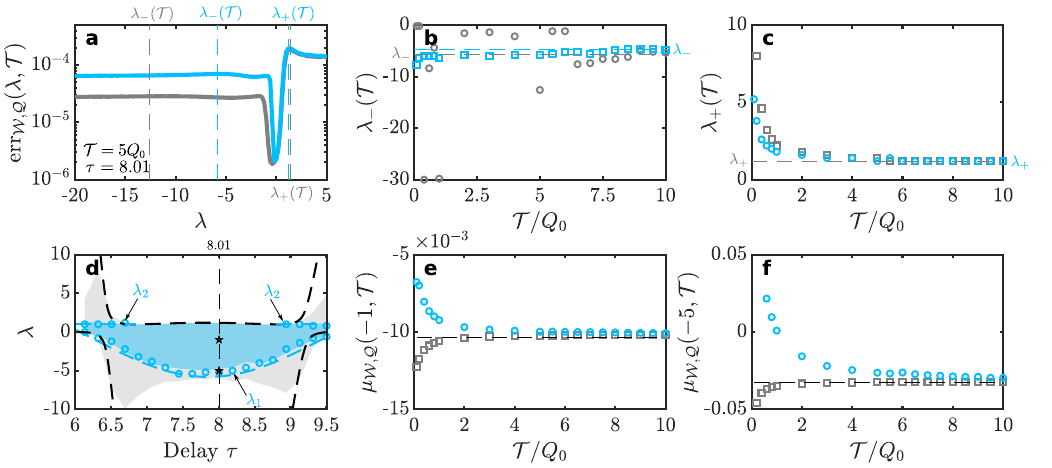}
	\caption{Same plots as in Fig.~2 of the main text for the delay $\tau=8.01$.
	}
	\label{Fig:linear_tau2}
\end{figure*}

\begin{figure*}[h!]	\includegraphics[width=\linewidth]{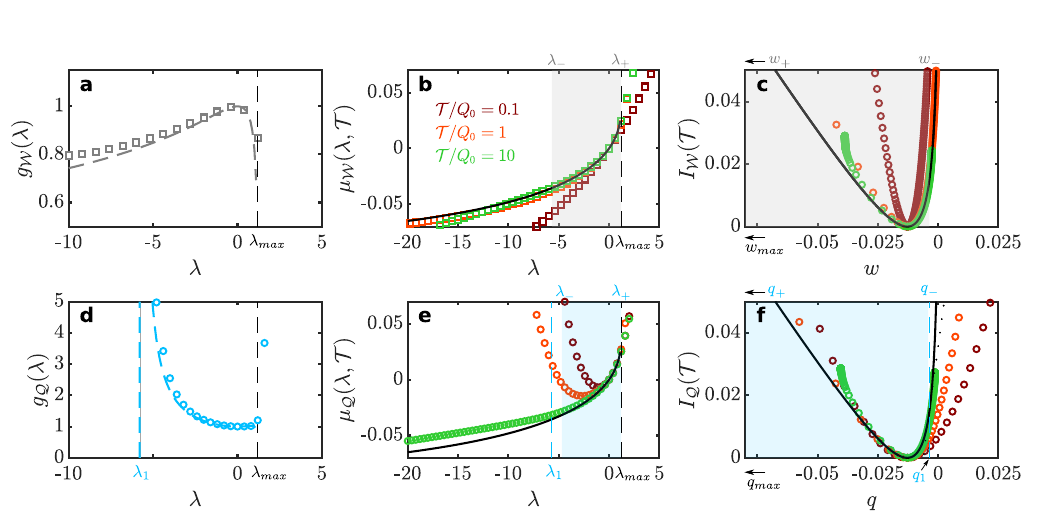}
	\caption{Same plots as in Fig.~3 of the main text for the delay $\tau=8.01$.
	}
	\label{Fig:ldf_tau2}
\end{figure*}

\subsection{d. Influence of the number $N$ of trajectories  on the convergence region}

In this section,  we analyze the role of the finite  number $N$ of trajectories  on the boundaries of the convergence regions. We choose the same delay as in the main text, $\tau=7.07$, and  vary the number of trajectories from $N=1000$ to $N=5 \cdot 10^6$. In Fig.~\ref{Fig:cvg_N} we show the behavior of $\lambda_\pm({\mathcal{T}})$ of work and heat versus the trajectory length and for various numbers of trajectories. The curves $\lambda_+({\mathcal{T}})$ all converge towards the same value $\lambda_+$ which coincides with $\lambda_{\max}$, the  boundary of the asymptotic domain of definition of the scaled cumulant generating function $\mu(\lambda)$ (panels c and d). We observe  in this case that $\lambda_+$ does not depend on the number of trajectories, as already noted in Ref.~\cite{S_roh15} for the case of exponential distributions. For $\lambda_-(\mathcal{T})$ the situation is different: while  the value of $\lambda_-$ of the heat is independent of $N$ when it approaches the singularity  $\lambda_1$ of the prefactor  (panel b), the value of $\lambda_-$ of the work depends on  the number $N$ of trajectories (panel a). This behavior is due to the absence of both a singularity of the prefactor and a boundary  of the asymptotic domain of definition of the scaled cumulant generating function $\mu(\lambda)$ in this instance.

\begin{figure*}[h!]	\includegraphics[width=0.62\linewidth]{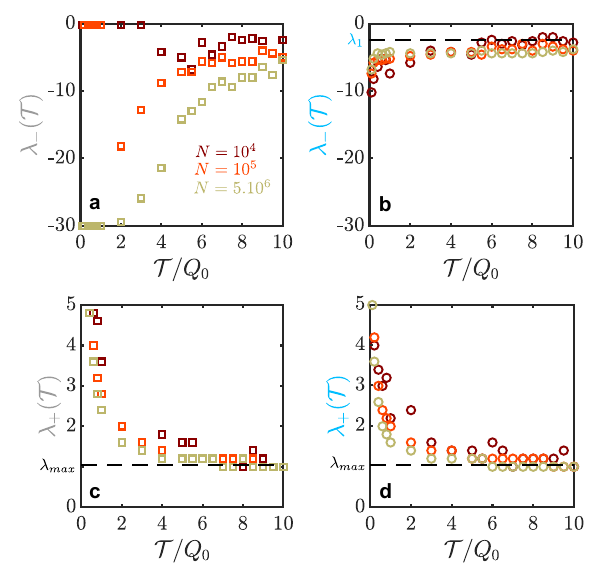}
	\caption{Convergence borders $\lambda_\pm(\mathcal{T})$ for work (a,c) and heat (b,d) as a function of the duration of the trajectory, for various number of trajectories, for the feedback delay $\tau=7.07$. In b) the horizontal dashed line indicate the position of the singularity of the prefactor $\lambda_1$. In (c,d) the horizontal dashed line indicates the limit $\lambda_{\max}$ of the asymptotic domain of definition of the scale cumulant generating function $\mu(\lambda)$. 
	}
	\label{Fig:cvg_N}
\end{figure*}

\subsection{e. Error estimates for the scaled cumulant generating function}

The error estimates on the scaled cumulant generating function are obtained using the block averaging method~\cite{S_zuc02}. We first randomly reshuffle a set of $N$ values of the averaged quantity $a=(a_1,a_2,...a_N)$, where $a$ stands for  work or heat, and reshape it into $N_b$ blocks containing $b=N/N_b$ values. In each block we calculate the scaled cumulant generating function as 
\begin{equation}
\mu_\mathcal{A}(\lambda,\mathcal{T},b)=\frac{1}{\mathcal{T}}\frac{1}{b}\ln\left(\sum_{i=1}^{b}e^{-\lambda\mathcal{A}_i}\right).
\end{equation}

The error on the scale cumulant generating function is then estimated as the standard error of the mean over the $N_b$ blocks:
\begin{equation}
\text{err}(\mu_\mathcal{A})=\sigma/\sqrt{N_b}
\end{equation}
with $\sigma$ the standard deviation over the blocks. In Fig.~\ref{Fig:err_mu} we show the error estimate on the scaled cumulant generating function for work and heat corresponding to the panels (b) and (e) of Fig.3 of the main text. We have used $N=5.10^6$ trajectories and $N_b=1000$ blocks. For $\lambda>0$, the errors on the scaled cumulant generating function for both work and heat increases rapidly as soon as $\lambda>\lambda_\text{max}$, in line with the result from Ref.~\cite{S_roh15} for exponential distributions. For $\lambda<0$, however, the situation is different. For small trajectory length, the error on the scaled cumulant generating function for heat is strongly affected by the presence of the singularity of the prefactor at $\lambda=\lambda_1$, which is not the case for work. For longer trajectory length, the effect of the singularity is suppressed since the statistics is not dictated anymore by the prefactor.          

\begin{figure*}[h!]	\includegraphics[width=0.8\linewidth]{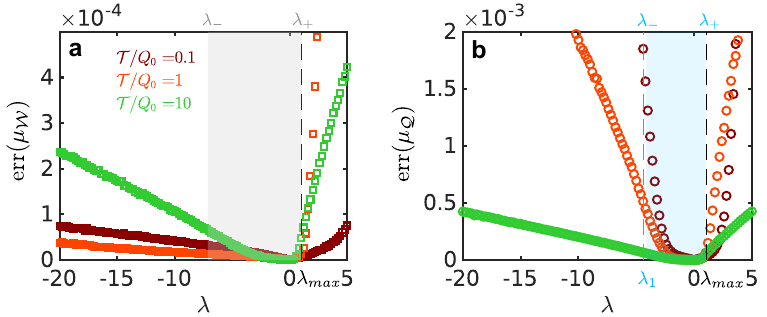}
	\caption{Statistical error of the scaled cumulant generating function for work (a) and heat (b), for various trajectory lengths and for a delay $\tau=7.07$. Here we have used $N=5.10^6$ trajectories and $N_b=1000$ blocks to estimate the error. Shaded areas represent the convergence borders $\lambda_\pm$ (see main text). Vertical black dashed lines indicates the limit $\lambda_\text{max}$ of the asymptotic domain of definition of $\mu(\lambda)$ and the vertical blue dashed line in (b) highlights the position of the singularity $\lambda_1$ of the prefactor for heat.     
	}
	\label{Fig:err_mu}
\end{figure*}


\begin{thebibliography}{99}
\bibitem{ell85} R. S. Ellis, \textit{Entropy, Large Deviations, and Statistical Mechanics}, (Springer, Berlin, 1985).
\bibitem{deu89} J. D. Deuschel and D. W. Stroock, \textit{Large Deviations}, (Academic Press, Boston, 1989).
\bibitem{dem98} A. Dembo and O. Zeitouni, \textit{Large Deviations Techniques and Applications}, (Springer, Berlin, 1998).
\bibitem{gia06} C. Giardin\`a, J. Kurchan, and L. Peliti, Direct Evaluation of Large-Deviation Functions, Phys. Rev. Lett. \textbf{96}, 120603 (2006).
\bibitem{rag18} F. Ragone, J. Wouters, and F. Bouchet, Computation of extreme heat waves in climate models using a large deviation algorithm, Proc. Natl. Acad. Sci. USA \textbf{24}, 15 (2018).

\bibitem{fri15} P. K. Friz, J. Gatheral, A. Gulisashvili, A. Jacquier, and J. Teichmann, \textit{Large Deviations and Asymptotic Methods in Finance}, (Springer, Berlin, 2015).
\bibitem{buc90} J. A. Bucklew, \textit{Large Deviation Techniques in Decision, Simulation and Estimation} (Wiley, New York, 1990).
\bibitem{wei95}  A. Weiss and A. Shwartz, \textit{Large Deviations for Performance Analysis}, (Chapman and Hall, London, 1995).


\bibitem{bre14} P. C. Bressloff, \textit{Stochastic Processes in Cell Biology}, (Springer, Berlin, 2014).
\bibitem{vul14} A. Vulpiani, F. Cecconi, M. Cencini, A. Puglisi and D. Vergni, \textit{Large Deviations in Physics}, (Springer, Berlin, 2014).
\bibitem{oon89} Y. Oono, {Large Deviation and Statistical Physics}, Prog. Theor. Phys.  \textbf{99}, 165 (1989). 
\bibitem{tou09} H. Touchette, The large deviation approach to statistical mechanics, Phys. Rep. \textbf{478}, 1 (2009).

\bibitem{bon09} M. Bonaldi, L. Conti, P. DeGregorio, L. Rondoni, G. Vedovato, A. Vinante, M. Bignotto, M. Cerdonio, P. Falferi, N. Liguori, S. Longo, R. Mezzena, A. Ortolan, G. A. Prodi, F. Salemi, L. Taffarello, S. Vitale, and J. P. Zendri, Nonequilibrium Steady-State Fluctuations in Actively Cooled Resonators, Phys. Rev. Lett. \textbf{103}, 010601 (2009).
\bibitem{kum11} N. Kumar, S. Ramaswamy, and A.K. Sood, Symmetry properties of the large-deviation function of the velocity of a self-propelled polar particle, Phys. Rev. Lett. \textbf{106}, 118001 (2011)


\bibitem{roh15} C. M. Rohwer, F. Angeletti, and H. Touchette, Convergence of large-deviation estimators, Phys. Rev. E \textbf{92}, 052104 (2015).
\bibitem{nem17} T. Nemoto, E. Guevara Hidalgo, and V. Lecomte, Finite-time and finite-size scalings in the evaluation of large-deviation functions: Analytical study using a birth-death process, Phys. Rev. E \textbf{95}, 012102 (2017).
\bibitem{hid17} E. Guevara Hidalgo, T. Nemoto, and V. Lecomte, Finite-time and finite-size scalings in the evaluation of large-deviation functions: Numerical approach in continuous time, Phys. Rev. E \textbf{95}, 062134 (2017).
\bibitem{whi18} S. Whitelam, Sampling rare fluctuations of discrete-time Markov chains, Phys. Rev. E \textbf{97}, 032122 (2018).
\bibitem{rag20} F. Ragone and F. Bouchet, Computation of extreme values of time averaged observables in climate models with large deviation techniques, J.  Stat.  Phys. \textbf{179}, 1637 (2020).


\bibitem{muz08}  J.-F. Muzy, E. Bacry, R. Baile, and P. Poggi, Uncovering latent singularities from multifractal scaling laws in mixed asymptotic regime. Application to turbulence, Europhys. Lett. \textbf{82}, 60007 (2008).
\bibitem{bac10} E. Bacry, A. Gloter, M. Hoffmann, and J.-F. Muzy, Multifractal analysis in a mixed asymptotic framework, Ann. Appl. Prob. \textbf{20}, 1729 (2010).

\bibitem{ber11}  L. Berthier and G. Biroli, Theoretical perspective on the glass transition and amorphous materials, Rev. Mod. Phys. \textbf{83}, 587 (2011).

\bibitem{gor03} J. Gore, F. Ritort, and C. Bustamante, Bias and error in
estimates of equilibrium free-energy differences from nonequi-
librium measurements, Proc. Natl. Acad. Sci. USA \textbf{100}, 12564 (2003).
\bibitem{jar06} C. Jarzynski, Rare events and the convergence of exponentially
averaged work values, Phys. Rev. E \textbf{73}, 046105 (2006).
\bibitem{cog23} F. Coghi, L. Buffoni and S. Gherardini, Convergence of the integral fluctuation theorem
estimator for nonequilibrium Markov systems, J. Stat. Mech. (2023) 063201.

\bibitem{sei12} U. Seifert, Stochastic thermodynamics, fluctuation theorems, and molecular machines, Rep. Prog. Phys. \textbf{75}, 126001 (2012).
 

\bibitem{gie18} J. Gieseler and J. Miller, Levitated Nanoparticles for Microscopic Thermodynamics--A Review, Entropy \textbf{20}, 326 (2018).
\bibitem{gie14} J. Gieseler, R. Quidant, C. Dellago, and L. Novotny,
Dynamic relaxation of a levitated nanoparticle from a non-equilibrium steady state, Nature Nanotechnol. \textbf{9}, 358 (2014).
\bibitem{hoa18} T. M. Hoang, R. Pan, J. Ahn, J. Bang, H. T. Quan, and T. Li,
Experimental Test of the Differential Fluctuation Theorem and a Generalized Jarzynski Equality for Arbitrary Initial
States, Phys. Rev. Lett. \textbf{120}, 080602 (2018).
\bibitem{deb20} M. Debiossac, D. Grass, J. J. Alonso, E. Lutz, and N. Kiesel, Thermodynamics of continuous non-Markovian feedback control, Nature Commun. \textbf{11}, 1360 (2020).
\bibitem{gon21} C. Gonzalez-Ballestero, M. Aspelmeyer, L. Novotny, R. Quidant, and O. Romero-Isart,  Levitodynamics: Levitation and control of microscopic objects in vacuum,  Science \textbf{374}, 168 (2021).
\bibitem{deb22} M. Debiossac, M. L. Rosinberg, E. Lutz, and N. Kiesel, Non-Markovian Feedback Control and Acausality: An Experimental Study,
Phys. Rev. Lett. \textbf{128}, 200601 (2022).



\bibitem{ytreberg2004} F. M. Ytreberg and D. M. Zuckerman, Efficient use of nonequilibrium measurement to estimate free energy differences for molecular systems, J. Comput. Chem. \textbf{25} 1749-1759 (2004).
\bibitem{duffy05} K. Duffy and A. P. Metcalfe, The large deviations of estimating rate functions, J. Appl. Prob. \textbf{42}, 267274 (2005).

\bibitem{ros17} M. L. Rosinberg, G. Tarjus, and T. Munakata, 
Stochastic thermodynamics of Langevin systems under time-delayed feedback control. II. Nonequilibrium steady-state fluctuations,
Phys. Rev. E \textbf{95}, 022123 (2017).


\bibitem{gra16} D. Grass, J.  Fesel, S. G. Hofer, N.  Kiesel and M. Aspelmeyer,  Optical trapping and control of nanoparticles inside evacuated hollow core photonic crystal fibers, Appl. Phys. Lett. \textbf{108}, 221103 (2016).

\bibitem{zon03} R. van Zon and E. G. D. Cohen, Extension of the Fluctuation Theorem Phys. Rev. Lett. \textbf{91}, 110601
(2003). 

\bibitem{vis06} P. Visco, Work fluctuations for a Brownian particle between two thermostats, J. Stat. Mech.  \textbf{P06006} (2006).
\bibitem{bai06} M. Baiesi, T. Jacobs, C. Maes, and N. S. Skantzos, Fluctuation symmetries for work and heat, Phys. Rev. E \textbf{74}, 021111 (2006).
\bibitem{noh12} J. D. Noh and J.-M. Park, Fluctuation relation for heat, Phys. Rev. Lett \textbf{108}, 240603 (2012).


\bibitem{kim14} K. Kim, C. Kwon, and H. Park, Heat fluctuations and initial ensembles, Phys. Rev. E \textbf{90}, 032117 (2014).
\bibitem{far02} J. Farago, Injected power fluctuations in Langevin equation, J. Stat. Phys. \textbf{107}, 781 (2002).
\bibitem{far04} J. Farago, Power fluctuations in stochastic models of dissipative systems, Physica A \textbf{331}, 69 (2004).
\bibitem{sab11}  S. Sabhapandit, Work fluctuations for a harmonic oscillator driven by an external random force, Euro. Phys. Lett. \textbf{96}, 20005 (2011).
\bibitem{sab12}  S. Sabhapandit, Heat and work fluctuations for a harmonic oscillator, Phys. Rev. E. \textbf{85}, 021108 (2012).
\bibitem{pug06} A. Puglisi, L. Rondoni, and A. Vulpiani, Relevance of initial and final conditions for the fluctuation relation in Markov processes, J. Stat. Mech. \textbf{P08010} (2006).
\bibitem{har06} R. J. Harris, A. R\'akos, and G. M. Sch\"utz, Breakdown of Gallavotti-Cohen symmetry for
stochastic dynamics, Europhys. Lett. \textbf{75}, 227 (2006).
\bibitem{nem12} T. Nemoto, Zon-Cohen singularity and negative inverse temperature in a trapped-particle limit, Phys. Rev. E \textbf{85}, 061124 (2012).

\end{thebibliography}

\begin{thebibliography}{11}
	
	
\bibitem{S_ros17} M. L. Rosinberg, G. Tarjus, and T. Munakata, 
Stochastic thermodynamics of Langevin systems under time-delayed feedback control. II. Nonequilibrium steady-state fluctuations, Phys. Rev. E \textbf{95}, 022123 (2017).

\bibitem{S_roh15} C. M. Rohwer, F. Angeletti, and H. Touchette, Convergence of large-deviation estimators, Phys. Rev. E \textbf{92}, 052104 (2015).

\bibitem{S_zuc02} Zuckerman, Daniel M., and Thomas B. Woolf, Overcoming finite-sampling errors in fast-switching free-energy estimates: extrapolative analysis of a molecular system. Chem. Phys. Lett. \textbf{351.5-6}, 445-453 (2002).
	
	
\end{thebibliography}
\end{document}